\documentclass{article}
\usepackage{authblk}
\usepackage{geometry}
\usepackage{graphicx}
\usepackage{caption}
\usepackage{xcolor}
\usepackage{float}
\usepackage{soul}
\graphicspath{Images}
\usepackage{amsmath, amssymb}
\usepackage{hyperref}
\usepackage{cancel}
\usepackage[fleqn]{mathtools}

\usepackage{cancel}
\usepackage{chngcntr}
\counterwithin*{equation}{section}
\counterwithin*{equation}{subsection}
\renewcommand{\theequation}{\mbox{\arabic{section}.\arabic{equation}}}
\renewcommand{\theequation}{\mbox{\arabic{section}.\arabic{subsection}.\arabic{equation}}}
\usepackage [english]{babel}
\usepackage [autostyle, english = american]{csquotes}
\MakeOuterQuote{"}
\usepackage{relsize}
\usepackage{multirow}
\usepackage{bigints}
\usepackage{amsmath, amssymb, physics, bm}
\usepackage{hyperref}
\usepackage{amsthm}
\theoremstyle{remark}

\usepackage{tikz}
\usetikzlibrary{arrows.meta,calc}
\usepackage[toc,page]{appendix}

\usepackage{authblk}
\usepackage{orcidlink}
\usepackage{hyperref}
\usepackage{tabularx}
\usepackage{booktabs}

\title{Ermakov--Lewis Invariants in Stationary Bohm--Madelung Quantum Mechanics}

\author[1]{Anand Aruna Kumar\,\orcidlink{0000-0001-6148-2777}}

\affil[1]{Research Engineer, IBM Research, Albany, NY, USA\par
	\href{mailto:anand.aruna.kumar@ibm.com}{anand.aruna.kumar@ibm.com}}

\date{} 

\begin{document}
	\maketitle
	
	\begin{abstract}
		\noindent The Ermakov–Pinney equation and its associated invariant are shown to arise
		naturally in stationary quantum mechanics when the Schr\"{o}dinger equation is
		expressed in Bohm–Madelung form and the Hamiltonian is diagonal and separable.
		Under these conditions, the stationary continuity constraint induces a nonlinear
		amplitude equation of Ermakov–Pinney type in each degree of freedom, revealing a
		hidden invariant structure that is independent of whether the evolution
		parameter is time or space.
		
		By reformulating the separated stationary equations in Sturm–Liouville form and applying Liouville normalization, we demonstrate that the quantum potential is encoded as a curvature contribution of the self-adjoint operator rather than appearing as an additional dynamical term. This correspondence preserves the standard probabilistic predictions of quantum mechanics while yielding exact stationary Bohmian amplitudes and their associated invariants. The resulting invariant-based formulation provides stationary guiding fields and clarifies the ontological status of Bohmian amplitudes as geometrically encoded structures rather than auxiliary dynamical additions. The results further show that stationary constrained Bohm–Madelung systems naturally admit variational formulations whose extremals preserve the Ermakov–Lewis invariant.
	\end{abstract}

	{\bf Keywords:} Ermakov–Lewis invariant; Bohm–Madelung mechanics; stationary quantum states; Sturm–Liouville theory; continuity constraint; separability.

	\section{Introduction}
	
	\noindent The stationary Bohm--Madelung (BM) formulation is discussed in Holland’s \textit{The Quantum Theory of Motion} \cite{PeterH}, where the one-dimensional amplitude equation is noted to be formally related to an Ermakov-type differential equation. This observation, however, is not developed into a general invariant framework. In the present work, we show that Ermakov--Pinney (EP) equations and their invariants arise naturally and systematically in stationary quantum mechanics for diagonal, separable Hamiltonians.
	\vspace{0.5pc}
	
	\noindent The EP equation and its associated invariant originate in the nineteenth-century work of Ermakov \cite{Ermakov1880} and were rediscovered and systematized by Lewis and Riesenfeld in the context of time-dependent harmonic oscillators \cite{Lewis1967}. Subsequent mathematical developments clarified the nonlinear superposition principle and invariant structure of Ermakov systems, notably in the work of Reid and collaborators \cite{Reid1971,ReidRay1984}, including the concise analysis of Reid and Ray on nonlinear oscillator invariants \cite{RiedRayZAMM}.
	\vspace{0.5pc}
	
	\noindent Connections between Ermakov-type differential equations and quantum mechanics have appeared sporadically, often within trajectory-based or hydrodynamic formulations \cite{NassarAB1986, NassarAB2017}. Despite these contributions, explicit treatments of the associated invariant structure in stationary BM equations and spatially separable Hamiltonians remain limited.
	\vspace{0.5pc}
	
	\noindent The present work addresses this gap by demonstrating that invariant structures arise naturally in stationary quantum mechanics whenever the Hamiltonian is diagonal and separable, with the Ermakov--Pinney (EP) equation emerging as the amplitude-space counterpart of Hamilton--Jacobi (HJ) separability. In this setting, the spatial coordinate plays the role of an evolution parameter, and the invariant structure appears at the level of stationary amplitude dynamics rather than through explicit time dependence.
	\vspace{0.5pc}
	
	\noindent The EP equation
	\begin{equation}
		\rho'' + \Omega^2(x)\rho = \frac{k}{\rho^3}
		\label{EP}
	\end{equation}
	plays a central role in the theory of time-dependent oscillators \cite{Nassar1985, Nassar1986} and admits a conserved Ermakov--Lewis invariant (henceforth referred to as the ``EL invariant'') \cite{Mancas2014}. In contrast, stationary quantum mechanics is usually presented as an eigenvalue problem, with little emphasis on hidden EL invariants beyond the energy.
	\vspace{0.5pc}
	
	\noindent Ermakov--Lewis invariant structures in quantum mechanics have been discussed in earlier works, particularly by Reinisch~\cite{Reinisch1994} and Schuch~\cite{Schuch2008,Schuch2018} in Bohm--Madelung and Riccati formulations. The present work extends the formalism to separable coordinate systems and demonstrates its application by analytically solving the Born--Oppenheimer two-center potential.
	\vspace{0.5pc}
	
	\noindent We show that this distinction is largely historical rather than structural. When stationary quantum systems are expressed in the BM representation, the same mathematical structure underlying EL invariants emerges intrinsically from the stationary continuity constraint and the separability of the Hamiltonian. For diagonal and separable Hamiltonians, this structure leads systematically to EP equations governing the amplitude dynamics in each degree of freedom, together with conserved EL invariants.
	\vspace{0.5pc}
	
	\noindent The analysis is therefore restricted to stationary quantum systems admitting coordinate-wise separability and diagonal kinetic energy, for which the BM equations reduce naturally to independent Sturm--Liouville sectors. Situations involving non-separable Hamiltonians, angular-variable subtleties, or gauge-induced momentum couplings that obstruct complete self-adjoint reduction are not pursued here, as they lie outside the structural regime in which the invariant framework arises transparently.
	\vspace{0.5pc}
	
	\noindent Given the broad scope of the present work, spanning variational formulations, coordinate-separable structures, and nonlinear representations of quantum mechanics, it is possible that some related contributions have not been cited. Any such omissions are unintentional, and the authors welcome further references to relevant literature.
	\section{Stationary BM Formulation and EP Structure}
	
	Consider the stationary Schr\"{o}dinger equation
	\begin{equation}
		\hat{H}\psi = E\psi,
	\end{equation}
	with the wave function $\psi$ in polar decomposition
	\begin{equation}
		\psi(q_1,\ldots,q_n) = R(q_1,\ldots,q_n)\,e^{iS(q_1,\ldots,q_n)/\hbar}.
	\end{equation}
	
	\noindent We assume a diagonal Hamiltonian of the form
	\begin{equation}
		H = \sum_{i=1}^{n}\frac{p_i^2}{2m} + V(q_1,\ldots,q_n),
	\end{equation}
	and separability,
	\begin{equation}
		R = \prod_i R_i(q_i), \qquad S = \sum_i S_i(q_i).
	\end{equation}
	
	\noindent Substitution into the Schr\"{o}dinger equation yields a set of continuity equations
	and separated energy equations for each degree of freedom.
	
	\subsection{Continuity Equation and Constraint}
	
	The stationary continuity equation for the probability current reads
	\begin{equation}
		\nabla \cdot (|\psi|^2 \mathbf{p}) = 0 .
	\end{equation}
	
	\noindent For separable stationary configurations we consider the branch in which each coordinate sector carries an independently conserved stationary current. In this case the continuity equation reduces to a set of sectorwise conditions
	\begin{equation}
		\frac{d}{dq_i}\left(R_i^2 p_i\right) = 0 ,
	\end{equation}
	which integrate immediately to
	\begin{equation}
		p_i = \frac{C_i}{R_i^2},
		\label{bohmcanonical}
	\end{equation}
	where $C_i$ are constants representing the conserved stationary flux in each coordinate channel.
	
	\subsection{EP Structure and the EL Invariant}
	
	The separated energy equations are
	\begin{equation}
		\frac{p_i^2}{2m} + V_i(q_i) + Q_{B,i} = E_i,
		\qquad \sum_i E_i = E,
	\end{equation}
	where $Q_{B,i}$ denotes the $q_i$--sector contribution of the Bohm quantum potential.
	In general orthogonal coordinates this sector term contains a coordinate--measure
	(first-derivative) contribution; it is therefore most naturally represented after
	placing the separated equation in Sturm--Liouville (Liouville-normalized) form (see
	Subsec.~\ref{subsec:SL_encoding}).
	
	In particular, once the separated equation has been brought to Liouville normal
	form, the amplitude variable in that sector obeys an EP equation. Writing
	the Liouville-scaled sector amplitude as $\rho_i(q_i)$, substitution of
	Eq.~\eqref{bohmcanonical} into the quantum HJ equation yields
	\begin{equation}
		\rho_i''(q_i) + \Omega_i^2(q_i)\,\rho_i(q_i) = \frac{k_i}{\rho_i^3(q_i)},
		\label{eq:EPi}
	\end{equation}
	which is the EP equation in the $q_i$--sector. Here $k_i=C_i^2/\hbar^2$
	is fixed by the conserved stationary probability flux in the $q_i$-sector, with $C_i$
	arising directly from the separated continuity equation rather than from boundary conditions imposed on the wavefunction.
	\noindent Thus, diagonal and separable stationary BM systems admit
	EP dynamics in the Liouville-normalized amplitude variable. In addition to the differential equation itself, the coupled equations for $\rho$ and its linearly independent companion admit a first integral that is independent of the coordinate 
	$q$, defining an invariant quantity.
	Specifically, if $y_i(q_i)$ is any solution of the associated linear partner equation
	\begin{equation}
		y_i''(q_i)+\Omega_i^2(q_i)\,y_i(q_i)=0,
		\label{eq:linearpartner}
	\end{equation}
	
	\noindent and $\rho_i(q_i)$ satisfies Eq.~(2.2.2), then
	
	\begin{equation}
		\boxed{
			I_i=\frac{1}{2}\left[\big(\rho_i y_i'-\rho_i' y_i\big)^2+\frac{k_i\,y_i^2}{\rho_i^2}\right]
		}
		\label{eq:ELinv}
	\end{equation}
	is independent of $q_i$. Equation~\eqref{bohmcanonical} is the key link between stationary
	Bohmian continuity and Ermakov theory.
	
	\subsection{Sturm--Liouville encoding of the quantum potential}
	\label{subsec:SL_encoding}
	
	A useful operator--level perspective is that, for separable stationary problems,
	the \emph{same} second--order differential operator simultaneously encodes\\
	(i) the linear Schr\"{o}dinger/Sturm--Liouville eigenproblem and\\
	(ii) the curvature term that appears in the Bohm quantum potential after polar decomposition.\\

	\noindent In particular, once a separated equation is written in self--adjoint Sturm--Liouville form,
	\begin{equation}
		\frac{d}{dq}\!\left(s(q)\frac{dX}{dq}\right)+s(q)\,Q_{\mathrm{SL}}(q)\,X(q)=0,
		\label{eq:SL_form_maintext}
	\end{equation}
	the Liouville substitution $X(q)=\psi(q)/\sqrt{s(q)}$ removes the first--derivative term and yields,
	\begin{equation}
		\psi''(q)+\Omega^2(q)\,\psi(q)=0,
		\qquad
		\Omega^2(q)=Q_{\mathrm{SL}}(q)-\frac12(\ln s)''-\frac14\big[(\ln s)'\big]^2.
		\label{eq:Omega_from_SL_maintext}
	\end{equation}
	
	\noindent In Liouville-normalized variables, the Bohm quantum potential is represented by the curvature contribution induced by Liouville normalization of the self-adjoint Sturm--Liouville operator, rather than appearing as an additional dynamical term.
	\vspace{0.5pc}
	
	\noindent The details of this normalization, including the cancellation of coordinate--measure
	terms in the Bohm quantum potential and the resulting diagonal (normal) form, are
	given in {\bf Appendices~A and~B}. The same $\Omega^2(q)$ is precisely the coefficient that appears in the EP
	amplitude equation obtained from the stationary Bohm continuity constraint, cf.\
	Eq.~\eqref{eq:EPi} (with linear partner \eqref{eq:linearpartner}).
	\vspace{0.5pc}
	
	\noindent Importantly, the Bohm quantum potential is defined from the physical amplitude
	$R$ as $Q_B=-(\hbar^2/2m)(\nabla^2R/R)$. However, once the separated sector is placed in
	Liouville normal form, the coordinate--measure contribution is absorbed into the operator,
	and the remaining \emph{sector curvature} term is encoded by the normal-form ratio
	$-(\hbar^2/2m)\,\psi''/\psi$. In this sense, no additional structure is introduced:
	the ``quantum potential curvature'' is already present in the Sturm--Liouville operator
	when written in normal form.
	\vspace{0.5pc}
	
	\noindent Before turning to specific potentials, it is useful to emphasize the general
	content of the correspondence established above. For stationary problems with
	separable Hamiltonians, the BM equations may be rewritten in a
	self-adjoint Sturm--Liouville form, in which the curvature term associated with
	the Bohm quantum potential is absorbed into the effective operator through
	Liouville normalization. The resulting amplitude equation is of EP
	type, and admits a conserved EL invariant, revealing a hidden invariant structure governing stationary amplitudes without altering the physical predictions of quantum mechanics.
	\vspace{0.5pc}
	
	\noindent At the level of the stationary Schr\"{o}dinger equation, the Sturm–Liouville formulation is conventionally employed to classify admissible eigenfunctions through boundary conditions for a given potential. In the BM formulation, however, the same self-adjoint structure acquires an additional role: the stationary continuity equation imposes global constraints on probability currents, promoting the amplitude and momentum fields to explicitly constrained configuration-space quantities. It is this interplay between self-adjointness and continuity—rather than the eigenvalue problem alone—that gives rise to the EP structure and its associated EL invariant, and enables the analytic construction of exact stationary Bohmian guiding fields whenever the underlying Sturm–Liouville problem is solvable.
	
	\paragraph{Implication for trajectories (stationary fields).}
	In stationary Bohmian mechanics the velocity field is
	\(v(x)=\dot{x}=p(x)/m=S'(x)/m\).
	For the separable stationary class considered here, $p(x)=C/R^2(x)$ is obtained
	algebraically from the continuity equation once the amplitude $R$ is known.
	Thus the trajectory equation reduces to a first--order quadrature
	$(\dot{x}={C}/{mR^2(x)})$
	(with analogous componentwise relations in separable multi--dimensional cases),
	so that the guiding field can be constructed analytically without resorting to
	stochastic sampling or numerical trajectory reconstruction. We emphasize that
	this does not add new physics beyond standard stationary quantum mechanics; it
	reorganizes the solution in amplitude--phase variables and makes the associated
	invariant structure explicit.

	\section{Canonical One-Dimensional Examples}
	
	The following examples illustrate how the EP structure manifests
	for the three canonical one-dimensional stationary quantum systems: the free
	particle, the harmonic oscillator, and the Coulomb potential.
	
	\subsection{Free particle: Bohm continuity \texorpdfstring{$\rightarrow$}{->} Ermakov amplitude and invariant}
	
	\label{subsec:free}
	
	For the free particle $V(x)=0$ and $E=\hbar^2k_0^2/(2m)$, hence
	\begin{equation}
		\Omega^2(x)=k_0^2.
	\end{equation}
	The Bohmian amplitude therefore satisfies the constant-frequency EP equation. For Cartesian coordinates the Sturm--Liouville weight is $s=1$, hence the physical amplitude and Liouville amplitude coincide ($R=\rho$) and the equation is
	
	\begin{equation}
		\rho''+k_0^2 \rho=\frac{k}{\rho^3},
		\qquad k=\frac{C^2}{\hbar^2}.
		\label{eq:EPfree}
	\end{equation}
	A particularly transparent solution is the constant-amplitude solution
	$\rho(x)=\rho_0$, for which \eqref{eq:EPfree} gives
	\begin{equation}
		k_0^2 \rho_0=\frac{k}{\rho_0^3}
		\quad\Rightarrow\quad
		\rho_0^4=\frac{k}{k_0^2}
		=\frac{C^2}{\hbar^2 k_0^2}
		\quad\Rightarrow\quad
		\rho_0=\left(\frac{|C|}{\hbar k_0}\right)^{1/2}.
	\end{equation}
	The Bohmian momentum $p=S'$ then becomes
	\begin{equation}
		p(x)=\frac{C}{\rho_0^2}=\mathrm{sgn}(C)\,\hbar k_0,
	\end{equation}
	recovering the constant momentum of the plane wave. The corresponding
	probability density is $|\psi|^2=\rho_0^2=\mathrm{const}$, consistent with standard
	quantum mechanics for scattering states. This constant-amplitude branch corresponds to delocalized scattering states, illustrating how the EL invariant naturally distinguishes bound and open sectors through the value of the conserved flux constant $C$.
	\vspace{0.5pc}
	
	\noindent To connect explicitly with the linear partner equation, note that the associated
	linear equation $y''+k_0^2 y=0$ admits the independent solutions
	\begin{equation}
		y_1(x)=\cos(k_0 x),\qquad y_2(x)=\sin(k_0 x),
		\qquad W=y_1y_2'-y_1'y_2=k_0.
	\end{equation}
	Hence the general Ermakov amplitude can be written as
	\begin{equation}
		\rho^2(x)=A\cos^2(k_0 x)+B\sin^2(k_0 x)+2D\sin(k_0 x)\cos(k_0 x),
		\qquad AB-D^2=\frac{k}{k_0^2}.
	\end{equation}
	\noindent The constant-amplitude solution corresponds to the symmetric choice
	$A=B=\rho_0^2$ and $D=0$, which satisfies $AB-D^2=\rho_0^4=k/k_0^2$.
	\vspace{0.5pc}
	
	\noindent Finally, for any linear solution $y$ and the Ermakov solution $\rho$, the invariant
	\eqref{eq:ELinv} is constant in $x$.
	
	\subsection{Harmonic oscillator: Weber basis and Bohmian Ermakov amplitude}
	
	For the one-dimensional harmonic oscillator with
	\(
	V(x)=\frac12 m\omega^2 x^2,
	\)
	the stationary BM reduction yields the EP equation for
	the amplitude \(\rho(x)\),
	\begin{equation}
		\rho''(x)+\left(\frac{2mE}{\hbar^2}-\frac{m^2\omega^2}{\hbar^2}x^2\right)\rho(x)
		=\frac{C^2}{\hbar^2}\frac{1}{\rho^3(x)}.
		\label{eq:EP_HO}
	\end{equation}
	The associated linear partner equation is obtained by setting \(C=0\):
	\begin{equation}
		y''(x)+\left(\frac{2mE}{\hbar^2}-\frac{m^2\omega^2}{\hbar^2}x^2\right)y(x)=0.
		\label{eq:HO_linear_partner}
	\end{equation}
	
	\noindent Introducing the dimensionless coordinate
	\begin{equation}
		\xi=\sqrt{\frac{m\omega}{\hbar}}\,x,
	\end{equation}
	Eq.~\eqref{eq:HO_linear_partner} reduces to the Weber equation
	\begin{equation}
		\frac{d^2 y}{d\xi^2}+\left(\nu+\frac12-\frac{\xi^2}{4}\right)y=0,
		\qquad\nu:=\frac{E}{\hbar\omega}-\frac12.
		\label{eq:Weber}
	\end{equation}
	A natural independent basis is provided by parabolic cylinder functions,
	\begin{equation}
		y_1(\xi)=D_\nu(\xi),\qquad y_2(\xi)=D_\nu(-\xi).
		\label{eq:Weber_basis}
	\end{equation}
	
	\noindent The general Bohmian amplitude solving the nonlinear EP equation
	\eqref{eq:EP_HO} can then be written in EP form using the same basis:
	\begin{equation}
		\rho^2(x)=A\,y_1^2(\xi)+B\,y_2^2(\xi)+2D\,y_1(\xi)y_2(\xi),
		\qquad	AB-D^2=\frac{k}{W^2},
		\label{eq:HO_Pinney}
	\end{equation}
	where \(W=y_1y_2'-y_1'y_2\) is the (constant) Wronskian and
	\(k=C^2/\hbar^2\). The  invariant
	defined in Eq.~\eqref{eq:ELinv} is therefore constant along \(x\) for any choice
	of \(y\) solving the Weber equation and \(\rho\) given by \eqref{eq:HO_Pinney}.
	
	\paragraph{Why the Weber Basis?}
	
	\noindent The structural aspect of solving the second order differential equation to establish the EL invariant is associated with the choice of co-ordinate basis. Naturally, for the harmonic oscillator, the linear partner equation associated with the EP system is the Weber equation, whose independent solutions
	are given by the parabolic cylinder functions \(D_\nu(\xi)\) and
	\(D_\nu(-\xi)\). In contrast to the conventional treatment, where the harmonic oscillator is
	formulated directly in terms of Hermite functions after imposing
	normalizability, the EP framework retains access to the full space
	of independent solutions prior to the application of boundary conditions.
	Accordingly, the invariant quadratic combination satisfies \eqref{eq:HO_Pinney}.
	\vspace{0.5pc}
	
	\noindent Here the Wronskian $W$ ensures linear independence, while the
	invariant guarantees consistency of the amplitude construction.
	Boundary and normalization conditions then select particular invariant
	combinations, rather than eliminating solutions at the outset. For discrete
	values $\nu=n$, the Weber functions reduce to Hermite functions,
	\(D_n(\xi)\propto e^{-\xi^2/4}H_n(\xi/\sqrt{2})\) \cite{Arfken}, recovering the standard
	harmonic-oscillator spectrum.
	\vspace{0.5pc}
	
	\subsection{Coulomb potential}
	
	For the one-dimensional Coulomb potential $V(x)=-\alpha/x$ on the half-line,
	the stationary equation reduces to Whittaker’s equation, which also arises from separation of the Helmholtz equation in paraboloidal coordinates.
	A fundamental solution pair is given by \(M_{\kappa,1/2}(2\lambda x)\) and \(W_{\kappa,1/2}(2\lambda x)\),
	leading to the general Ermakov amplitude
	\[
	\rho^2(x)=A M_{\kappa,1/2}^2 + B W_{\kappa,1/2}^2 + 2D M_{\kappa,1/2}W_{\kappa,1/2}.
	\]
	with \(AB-D^2=k/W^2\) given by the Wronskian relation associated with the EL invariant. Imposing the quantization condition
	\(\kappa=n+1\) reduces the Whittaker \(M\)-function to the familiar Laguerre form \cite{Arfken},
	\[
	M_{n+1,1/2}(z)\propto z e^{-z/2} L_n^{(1)}(z),
	\]
	recovering the textbook Coulomb bound states as a constrained Ermakov limit.

	\section{Well-Posedness of the Stationary Bohm–Ermakov Problem}

	In classical mechanics, Hamilton--Jacobi separability guarantees integrability
	of the phase function $S$. In Bohmian mechanics, this structure is mirrored:
	\begin{itemize}
		\item the phase $S_i$ satisfies HJ equations,
		\item the amplitude $\rho_i$ satisfies EP equations.
	\end{itemize}
	
	\noindent EL invariants therefore represent the amplitude-space dual of
	HJ separability.
	
	\subsection{ Relation to Hamilton--Jacobi Separability  }
	
	Each separated stationary equation reduces to a Helmholtz or Sturm--Liouville
	problem
	\begin{equation}
		y'' + \Omega^2(q)y = 0.
	\end{equation}
	
	\noindent Given two independent solutions $y_1,y_2$ with Wronskian $W$, the general
	Ermakov solution for the Liouville-normalized amplitude is
	\begin{equation}
		\rho^2 = A y_1^2 + B y_2^2 + 2D y_1 y_2,
		\qquad AB - D^2 = \frac{k}{W^2}.
	\end{equation}
	
	\noindent  Standard quantum mechanics typically restricts this structure through boundary conditions that select a single admissible solution.
	
	\vspace{0.5pc}
	
	\noindent EL invariants thus provide a smooth bridge between classical action
	conservation and quantum amplitude dynamics.
	\vspace{0.5pc}

	\subsection{Implications}
	
	This explicit coordinate-level analysis confirms that the appearance of
	EP dynamics in stationary quantum mechanics is a structural
	consequence of separability and diagonal kinetic energy, and not an artifact
	of a particular coordinate choice. Gauge-induced momentum couplings, such as those arising from magnetic vector potentials, generally destroy the diagonal structure required for complete Sturm--Liouville separability. However, when the gauge choice preserves the
	underlying coordinate symmetries (for example, the symmetric gauge in
	cylindrical geometry), the Hamiltonian can still be reduced to independent
	self-adjoint sectors. In such cases the BM reduction remains
	compatible with Liouville normalization and continues to admit EP
	invariants.

	\subsection{Well-Posedness Criteria and Invariant Constraints}
	
	The appearance of an EP equation in the stationary
	BM formulation does not alter the underlying eigenvalue
	problem, but reorganizes it into a form in which global constraints and
	integrability conditions are made explicit. In this framework, the
	stationary problem is posed as follows.
	\vspace{0.5pc}
	
	\noindent For a diagonal and separable Hamiltonian, separation of variables
	reduces the stationary Schr\"{o}dinger equation to a set of coordinate-wise
	Sturm--Liouville problems. The Bohm continuity equation then enforces a
	nonlinear constraint on the amplitude, yielding an associated
	EP equation for each degree of freedom. This equation exists
	independently of boundary conditions and admits a conserved
	EL invariant.
	\vspace{0.5pc}
	
	\noindent Physical solutions are selected not by modifying the differential
	equation itself, but by imposing global admissibility conditions on the
	invariant. These conditions include regularity, boundedness, and the
	presence or absence of stationary probability flux. In particular, the
	constant $k=C^2/\hbar^2$ appearing in the EP equation
	labels distinct solution sectors: confined or bound-state sectors are
	typically associated with vanishing stationary current ($C=0$), while
	open or scattering sectors naturally admit $C\neq0$.
	\vspace{0.5pc}
	
	\noindent Within each admissible sector, the amplitude may be expressed as a
	quadratic form in a fundamental pair of independent solutions of the
	associated linear Sturm--Liouville equation. Boundary and normalization
	requirements then restrict the allowed invariant combinations without
	excluding solutions a priori. In this way, the EL invariant
	organizes the solution space while preserving the standard stationary
	quantum-mechanical content of the problem.
	\vspace{0.5pc}
	
	\noindent This formulation clarifies that the Bohm quantum potential does not
	introduce additional dynamical degrees of freedom in the stationary
	case. Instead, it is encoded implicitly in the Sturm--Liouville geometry
	of the separated equations, with the EL invariant providing a
	compact representation of this structure.
	\vspace{0.5pc}
	
	\noindent A nontrivial illustration of this framework for a genuinely non-central potential is given in {\bf Appendix C}, where the two-center Coulomb problem separates in elliptic coordinates and admits explicit spatial EL invariants.
	
	\section{Conclusions}
	
	We have shown that Ermakov–Pinney equations and their associated invariants arise naturally in stationary quantum mechanics when the Schr\"{o}dinger equation is written in Bohm–Madelung form and the Hamiltonian is diagonal and separable. In this setting, the spatial coordinate plays the role of an evolution parameter, and the amplitude dynamics are governed by Sturm–Liouville operators whose Liouville-normal form induces the Ermakov structure.
	\vspace{0.5pc}
	
	\noindent This framework unifies a broad class of stationary problems—ranging from free motion and harmonic confinement to Coulombic and multi-center potentials—within a single invariant-based description, while preserving the standard probabilistic predictions of quantum mechanics. The Bohm quantum potential does not enter as an additional postulate; rather, it is encoded as the curvature contribution that emerges when the separated equations are written in Liouville-normalized Sturm–Liouville form.
	\vspace{0.5pc}
	
	\noindent From this perspective, the Ermakov--Lewis invariant may be interpreted as labeling invariant-preserving stationary flow configurations of the Bohmian guiding field, suggesting a constrained variational reinterpretation of the stationary problem without introducing additional dynamical structure.
	\vspace{0.5pc}
	
	\noindent Although the present analysis is restricted to separable Hamiltonians with diagonal kinetic terms, it provides a clear analytical route to constructing exact stationary Bohmian guiding fields without numerical trajectory simulations. Extensions beyond this class may be explored within the same invariant-based framework, but lie beyond the scope of the present work.
	
	\section*{Declaration of Interests}
	The research is an independently developed idea of the author, and declares no conflicts of interest with the employer, IBM Research, Albany, NY, USA.

	\section*{Acknowledgements}
	The author thanks his collaborators, S.K. Srivatsa and Rajesh Tengli, who co-authored on a related publication in connection to Bohmian stationary states. The author used generative AI tools to assist with cross-verification of algebraic expressions, consistency checks of equations, LaTeX formatting, and language editing. All scientific reasoning, derivations, interpretations, and conclusions are the sole responsibility of the author.

	\newpage
	
	\appendix
	\renewcommand{\thesection}{Appendix~\Alph{section}}
	\renewcommand{\theequation}{\Alph{section}.\arabic{equation}}
	
	\setcounter{equation}{0}
	
	\section{Coordinate Separation of the Bohm Quantum Potential}
	
	\label{AppendixA}
	
	This appendix provides the technical details underlying the separation of the
	Bohm quantum potential in stationary quantum mechanics when the Hamiltonian is
	diagonal and separable. The purpose is to demonstrate explicitly that the
	EP structure derived in the main text is not an artifact of a
	particular coordinate choice, but follows generally for orthogonal coordinate
	systems after suitable amplitude rescalings.
	
	\subsection*{General form of the quantum potential}
	
	The Bohm quantum potential is defined as
	\begin{equation}
		Q_B = -\frac{\hbar^2}{2m}\frac{\nabla^2 R}{R},
	\end{equation}
	where \(R\) is the amplitude in the polar decomposition
	\(\psi = R e^{iS/\hbar}\).
	\vspace{0.5pc}
	
	\noindent Assume separability of the amplitude,
	\begin{equation}
		R(q_1,\dots,q_n) = \prod_{i=1}^n R_i(q_i),
	\end{equation}
	in an orthogonal coordinate system \(\{q_i\}\) with diagonal metric
	coefficients \(h_i(q_i)\). 
	\vspace{0.5pc}
	
	\noindent Substituting the product form of \(R\) yields
	
	\begin{equation}
		\nabla^2 R
		=\frac{1}{\prod_{j=1}^n h_j}
		\sum_{i=1}^n\frac{\partial}{\partial q_i}
		\left(\frac{\prod_{j=1}^n h_j}{h_i^2}\,
		\frac{\partial R}{\partial q_i}\right),
	\end{equation}
	and hence
	\begin{equation}
		\frac{\nabla^2 R}{R}
		=\frac{1}{\prod_{j=1}^n h_j}
		\sum_{i=1}^n\frac{1}{R}\frac{\partial}{\partial q_i}
		\left(\frac{\prod_{j=1}^n h_j}{h_i^2}\,
		\frac{\partial R}{\partial q_i}\right).
		\label{eq:nabla2R_over_R_general}
	\end{equation}
	Assuming separability $R(q_1,\dots,q_n)=\prod_{k=1}^n R_k(q_k)$, one has
	$\partial_{q_i}R=R\,(R_i'/R_i)$, and therefore
	\begin{equation}
		\frac{\nabla^2 R}{R}
		=\frac{1}{\prod_{j=1}^n h_j}
		\sum_{i=1}^n\frac{\partial}{\partial q_i}
		\left(\frac{\prod_{j=1}^n h_j}{h_i^2}\,
		\frac{R_i'}{R_i}\right).
		\label{eq:nabla2R_over_R_general_sep}
	\end{equation}
	
	\noindent demonstrating that the quantum potential separates additively,
	\begin{equation}
		Q_B = \sum_{i=1}^n Q_{B,i}(q_i).
	\end{equation}
	
	\noindent Equivalently, expanding the derivative gives a form that isolates the usual
	second-derivative term and a ``measure'' contribution from the coordinate
	system:
	\begin{equation}
		\frac{\nabla^2 R}{R} =
		\sum_{i=1}^n\frac{1}{h_i^2}\frac{R_i''}{R_i} + \sum_{i=1}^n\frac{1}{h_i^2}
		\left[\frac{\partial}{\partial q_i}\ln\!\left(\frac{\prod_{j=1}^n h_j}{h_i^2}\right)\right]\frac{R_i' }{R_i}.
		\label{eq:nabla2R_over_R_expanded}
	\end{equation}
	\noindent This additive structure is essential for the emergence of independent
	EP equations for each degree of freedom.

	\paragraph{Additive structure and emergence of EP equations.}
	For diagonal and separable Hamiltonians in orthogonal coordinates, the Bohm
	quantum potential admits an additive decomposition,
	\begin{equation}
		Q_B = -\frac{\hbar^2}{2m}\frac{\nabla^2 R}{R}	= \sum_i Q_{B,i}(q_i),
	\end{equation}
	provided the amplitude factorizes as \(R=\prod_i R_i(q_i)\) and the coordinate
	operators are in Sturm--Liouville form. This additive structure is essential,
	as it allows each coordinate contribution to enter the quantum
	HJ equation independently.
	\vspace{0.5pc}
	
	\noindent At the same time, separation of the stationary continuity equation under the componentwise vanishing condition implies that each Bohmian momentum component satisfies
	\begin{equation}
		p_i(q_i)=\frac{C_i}{R_i^2(q_i)},
	\end{equation}
	where $C_i$ are separation constants representing the conserved stationary flux in the $q_i$ sector. Substitution into the quantum HJ equation then yields, for each
	degree of freedom, a nonlinear second-order equation of the EP
	type,
	\vspace{0.5pc}

	\begin{equation}
		R_i''(q_i)+P_i(q_i)\,R_i'(q_i)+\widetilde{\Omega}_i^2(q_i)\,R_i(q_i)
		=\frac{C_i^2}{\hbar^2}\,\frac{1}{R_i^3(q_i)},
		\label{eq:Ri_preLiouville_EP}
	\end{equation}
	where $P_i(q_i)$ is the coordinate-measure contribution (the same term that appears in
	\eqref{eq:nabla2R_over_R_expanded}. This equation is not in diagonal (normal) form.
	\vspace{0.5pc}
	
	\noindent To remove the first-derivative term, we apply the standard Liouville normalization
	\begin{equation}
		R_i(q_i)=\frac{\rho_i(q_i)}{\sqrt{s_i(q_i)}},
		\label{eq:Liouville_Ri_rhoi}
	\end{equation}
	with $s_i(q_i)$ the Sturm--Liouville weight for the $q_i$--sector. In terms of the
	Liouville-scaled amplitude $\rho_i$, the equation takes the EP form
	\begin{equation}
		\rho_i''(q_i)+\Omega_i^2(q_i)\,\rho_i(q_i)
		=\frac{C_i^2}{\hbar^2}\,\frac{1}{\rho_i^3(q_i)}.
		\label{eq:EP_rhoi}
	\end{equation}
	
	\noindent Thus, the EP structure arises directly from the combination of
	additive separability and the Bohm continuity constraint.

	\subsection*{General orthogonal-coordinate statement\\ (Sturm--Liouville $\to$ Liouville normal form)}
	
	The additive decomposition \eqref{eq:nabla2R_over_R_expanded} shows that each sector
	$q_i$ contributes a second-derivative term together with a coordinate-measure term
	proportional to $(R_i'/R_i)$.  For separable orthogonal systems this sector equation
	is naturally written in self-adjoint Sturm--Liouville form,
	\begin{equation}
		\label{eq:SL_sector_appendix}
		\frac{d}{dq_i}\!\left(s_i(q_i)\frac{dR_i}{dq_i}\right)
		+ s_i(q_i)\,\Omega_i^2(q_i)\,R_i(q_i)=0,
	\end{equation}
	where the (positive) weight $s_i$ is inherited from the metric prefactor
	(e.g. $s_i\propto (h_1h_2h_3)/h_i^2$ in $\mathbb{R}^3$).
	\vspace{0.5pc}
	
	\noindent Expanding \eqref{eq:SL_sector_appendix} gives the explicit first-derivative term,
	\begin{equation}
		R_i'' + \big(\ln s_i\big)'\,R_i' + \Omega_i^2\,R_i = 0,
		\label{eq:R_with_first_derivative}
	\end{equation}
	which is the origin of the ``measure'' contribution in \eqref{eq:nabla2R_over_R_expanded}.
	\vspace{0.5pc}
	
	\noindent The first-derivative term is removed identically by the Liouville normalization
	\begin{equation}
		\label{eq:Liouville_appendix}
		R_i(q_i)=\frac{\rho_i(q_i)}{\sqrt{s_i(q_i)}}.
	\end{equation}
	Substituting \eqref{eq:Liouville_appendix} into \eqref{eq:R_with_first_derivative} yields the diagonal normal form
	\begin{equation}
		\label{eq:rho_linear}
		\rho_i'' + \widetilde{\Omega}_i^2(q_i)\,\rho_i = 0,
		\qquad
		\widetilde{\Omega}_i^2
		= \Omega_i^2 -\frac12(\ln s_i)''-\frac14\big[(\ln s_i)'\big]^2 .
	\end{equation}
	
	\noindent Finally, separation of the stationary continuity equation implies
	$p_i=C_i/R_i^2$, and in terms of the Liouville-scaled amplitude $\rho_i$
	this yields the EP equation
	\begin{equation}
		\label{eq:rho_EP_appendix}
		\rho_i'' + \widetilde{\Omega}_i^2(q_i)\,\rho_i = \frac{C_i^2}{\hbar^2}\,\frac{1}{\rho_i^3}.
	\end{equation}
	Thus the EP structure is a consequence of (i) Sturm--Liouville self-adjointness,
	(ii) Liouville normalization, and (iii) the stationary Bohm continuity constraint,
	and is not specific to any particular coordinate choice.
	
	\subsection*{Cylindrical-coordinate scaling as an example}
	
	\noindent
	In cylindrical coordinates $(r,\phi,z)$ with scale factors
	$h_r=1$, $h_\phi=r$, and $h_z=1$, the radial amplitude equation
	appears in Sturm--Liouville form with weight $s_r=r$. Writing
	\begin{equation}
		R_r(r)=\frac{\rho_r(r)}{\sqrt{s_r(r)}}=\frac{\rho_r(r)}{\sqrt{r}},
	\end{equation}
	the first-derivative term originating from the coordinate measure
	cancels identically, yielding a second-order equation for $\rho_r$
	in Liouville normal form.
	\vspace{0.5pc}
	
	\noindent
	This illustrates the general mechanism: in any orthogonal separable
	coordinate system, separated amplitude equations inherit a
	Sturm--Liouville weight $s_i(q_i)$ determined by the metric.
	Liouville normalization,
	\begin{equation}
		R_i(q_i)=\frac{\rho_i(q_i)}{\sqrt{s_i(q_i)}},
	\end{equation}
	removes all measure-induced first derivatives and brings each sector
	to diagonal form. Substitution into the stationary Bohm
	HJ equation then yields the universal EP
	equation
	\begin{equation}
		\rho_i''(q_i)+\Omega_i^2(q_i)\rho_i(q_i)=\frac{k_i}{\rho_i^3(q_i)},
	\end{equation}
	showing that the invariant structure arises from coordinate geometry
	combined with the Bohm continuity constraint, rather than from any
	coordinate-specific property.
	
	\newpage
	
	\section{Coordinate Systems}
	\label{AppendixB}
	\setcounter{equation}{0}
	\renewcommand{\thesection}{Appendix~\Alph{section}}
	\renewcommand{\theequation}{\Alph{section}.\arabic{equation}}
	\subsection*{Spatial Separation and Sturm--Liouville Structure}
	
	Consider the Helmholtz equation
	\begin{equation}
		\nabla^2 \Psi + k^2 \Psi = 0
	\end{equation}
	in orthogonal coordinates $(q_1,q_2,q_3)$ with scale factors $h_i(q_1,q_2,q_3)$. The Laplacian takes the standard form
	\begin{equation}
		\nabla^2=\frac{1}{h_1 h_2 h_3}	\sum_{i=1}^3\frac{\partial}{\partial q_i}
		\left(\frac{h_1 h_2 h_3}{h_i^2} \frac{\partial}{\partial q_i}\right).
	\end{equation}
	
	\noindent For St\"{a}ckel-separable systems, separation of variables
	\[\Psi(q_1,q_2,q_3) = \prod_{i=1}^3 X_i(q_i)\]
	yields three ordinary differential equations of Sturm--Liouville type,
	\begin{equation}
		\frac{d}{dq_i}	\left(s_i(q_i)\frac{dX_i}{dq_i}	\right)	+ s_i(q_i) Q_{SL,i}(q_i) X_i = 0,
	\end{equation}
	where the weight function $s_i(q_i)$ is inherited from the geometric prefactor
	\begin{equation}
		s_i(q_i) \propto \left[ \frac{h_1 h_2 h_3}{h_i^2} \right] \text{ ($h_i$ depends on $q_i$)}
	\end{equation}
	
	\noindent This structure holds for all eleven orthogonal separable coordinate systems in $\mathbb{R}^3$. Thus, separation produces three independent Sturm--Liouville problems per coordinate system, which is the geometric quantum potential tabulated in Table~(\ref{Tab1}). The physical contribution to the effective frequency (defined explicitly below) may be written as
	\begin{equation}
		\label{eq:OmegaPhys}
		\Omega_{i,\mathrm{phys}}^2(q_i)
		=
		\frac{2m}{\hbar^2}\Big(E_i-V_i(q_i)\Big)
		-\frac{\kappa_i}{s_i^2(q_i)},
	\end{equation}
	where $V_i(q_i)$ denotes the separated potential contribution in the $q_i$--sector (including separation constants absorbed into $E_i$ as appropriate), and $\kappa_i$ is a nonnegative constant proportional to the square of the conserved stationary flux in that sector.
	In the zero--flux branch one has $\kappa_i=0$, in which case $\Omega_{i,\mathrm{phys}}^2=(2m/\hbar^2)(E_i-V_i)$.
	
	\subsection*{Liouville Normal Form and Ermakov Structure}
	
	Each Sturm--Liouville equation may be mapped to diagonal (normal) form via the Liouville transformation given by
	\begin{equation}
		X_i(q_i) = \frac{\psi_i(q_i)}{\sqrt{s_i(q_i)}}.
	\end{equation}
	The resulting equation is
	\begin{equation}
		\psi_i''(q_i) + \Omega_i^2(q_i)\,\psi_i(q_i) = 0,
	\end{equation}
	with effective frequency
	\begin{equation}
		\Omega_i^2(q_i)	=	Q_{SL,i}(q_i)-\frac12 (\ln s_i)''	-\frac14 \big[(\ln s_i)'\big]^2.
	\end{equation}
	Associated with this normal form is the EP equation (obtained from continuity equation)
	\begin{equation}
		\rho_i''(q_i) + \Omega_i^2(q_i)\rho_i(q_i)	=
		\frac{c_i}{\rho_i^3(q_i)},
	\end{equation}
	and the Ermakov--Lewis invariant
	\begin{equation}
		I_i	=	\frac12
		\left[	(\rho_i\psi_i' - \rho_i'\psi_i)^2	+	c_i\left(\frac{\psi_i}{\rho_i}\right)^2	\right].
	\end{equation}
	
	\noindent Thus, each separable coordinate contributes one EL invariant, yielding exactly three coordinate-wise invariants per orthogonal coordinate system. Cyclic coordinates correspond to trivial constant-frequency cases. To make explicit the respective roles of coordinate geometry and physical input in the effective frequency, it is convenient to separate $\Omega_i^2$ into geometric and physical parts.

	\paragraph{Physical versus geometric contributions in $\Omega_i^2$.}
	In the stationary BM formulation, separation together with the continuity constraint implies a conserved flux in each separated coordinate sector. Writing the Liouville-scaled separated amplitude as
	\begin{equation}
		X_i(q_i)=\frac{\psi_i(q_i)}{\sqrt{s_i(q_i)}},
	\end{equation}
	the Liouville normal form can be expressed as
	\begin{equation}
		\label{eq:liouvilleOmegaSplit}
		\psi_i''+\Omega_i^2(q_i)\,\psi_i=0,
		\qquad 
		\Omega_i^2(q_i)=\Omega_{i,\mathrm{geom}}^2(q_i)+\Omega_{i,\mathrm{phys}}^2(q_i) 
	\end{equation}
	The geometric part is induced solely by the Liouville scaling (after removal of the first derivative),
	\begin{equation}
		\label{eq:OmegaGeom}
		\begin{split}
			\Omega_{i,\mathrm{geom}}^2(q_i)
			&=
			-\frac{s_i''}{2s_i}
			+\frac{1}{4}\left(\frac{s_i'}{s_i}\right)^2, \\
			Q_i(q_i)
			&=
			-\frac{\hbar^2}{2m}\,\Omega_{i,\mathrm{geom}}^2(q_i)
			\;\text{(geometric quantum potential)} .
		\end{split}
	\end{equation}
	
	is tabulated in Table~(\ref{Tab1}). The remaining (physical) contribution may be written compactly as
	\begin{equation}
		\label{eq:OmegaPhysEV}
		\Omega_{i,\mathrm{phys}}^2(q_i)
		=
		\frac{2m}{\hbar^2}\Big(E_i-V_i(q_i)\Big)
		-\frac{\kappa_i}{s_i^2(q_i)},
	\end{equation}
	where $V_i(q_i)$ denotes the separated potential contribution in the $q_i$--sector (with separation constants absorbed into $E_i$ as appropriate), and $\kappa_i$ is a nonnegative constant proportional to the square of the conserved stationary flux in that sector. In the zero--flux branch one has $\kappa_i=0$, so that $\Omega_{i,\mathrm{phys}}^2=(2m/\hbar^2)(E_i-V_i)$.

	\subsection*{Separability and effective $\Omega$ for each coordinate system}

	Explicit Liouville scaling in eleven separable coordinate systems and the corresponding geometric quantum potential terms are summarized in Table~(\ref{Tab1}). For the ellipsoidal coordinate system, the Liouville scaling is trivial ($s_i=1$). However, depending on the roots of the associated parametric equation, the orthogonal coordinate surfaces may be ellipsoidal, hyperboloidal, or conical; these form the confocal quadric family summarized in Table~(\ref{Tab2}). Table~(\ref{Tab1}) summarizes standard coordinate systems, while Table~(\ref{Tab2}) covers ellipsoidal surfaces.

	\[\frac{d}{dq_i}\!\left(s_i(q_i)\frac{dX_i}{dq_i}\right)+\cdots=0,\]
	and the Liouville transformation
	\[X_i(q_i)=\frac{\psi_i(q_i)}{\sqrt{s_i(q_i)}}\]
	brings the equation to diagonal (Ermakov-ready) normal form
	$\psi_i''+\Omega_i^2(q_i)\psi_i=0$.

	
	\begin{table}[!h]
		\centering
		\footnotesize
		\setlength{\tabcolsep}{6pt}
		\renewcommand{\arraystretch}{1.1}
		
		\begin{tabularx}{\columnwidth}
			{l c c c >{\raggedright\arraybackslash}X}
			\toprule
			\textbf{System} &
			\textbf{$q_i$} &
			$\bm{s_i(q_i)}$ &
			\textbf{$X_i=\psi_i/\sqrt{s_i}$} &
			\textbf{Effective $\Omega^2(q_i)$} \\
			\midrule
			
			Cartesian $(x,y,z)$
			& $x,y,z$ & $1$ & $X_i=\psi_i$ & $k_i^2$ \\
			
			\midrule
			\multirow{3}{*}{Cylindrical}
			& $r$ & $r$ & $X_r=\psi_r/\sqrt{r}$ &
			$(k_0^2-k_z^2)-\dfrac{m_\theta^2-\tfrac14}{r^2}$ \\
			& $\theta$ & $1$ & $X_\theta=\psi_\theta$ & $m_\theta^2$ \\
			& $z$ & $1$ & $X_z=\psi_z$ & $k_z^2$ \\
			
			\midrule
			\multirow{3}{*}{Spherical}
			& $r$ & $r^2$ & $X_r=\psi_r/r$ &
			$k_0^2-\dfrac{l(l+1)}{r^2}$ \\
			& $\theta$ & $\sin\theta$ &
			$X_\theta=\psi_\theta/\sqrt{\sin\theta}$ &
			$l(l+1)+\tfrac14-\dfrac{m_\phi^2-\tfrac14}{\sin^2\theta}$ \\
			& $\phi$ & $1$ & $X_\phi=\psi_\phi$ & $m_\phi^2$ \\
			
			\midrule
			\multirow{3}{*}{3D Parabolic}
			& $u$ & $u$ & $U=\psi_u/\sqrt{u}$ &
			$k_0^2u^2+\lambda-\dfrac{m_u^2-\tfrac14}{u^2}$ \\
			& $v$ & $v$ & $V=\psi_v/\sqrt{v}$ &
			$k_0^2v^2-\lambda-\dfrac{m_v^2-\tfrac14}{v^2}$ \\
			& $\phi$ & $1$ & $\Phi=\psi_\phi$ & $m_\phi^2$ \\
			
			\midrule
			\multirow{1}{*}{Elliptic cyl.\,$^\dagger$}
			& $\mu,\nu$ &
			$a\sqrt{\sinh^2\mu+\sin^2\nu}$ &
			$X=\psi/\sqrt{s}$ &
			$Q_{SL}(q)-\dfrac{s''}{2s}+\dfrac{(s')^2}{4s^2}$ \\
			$^\dagger$: $q=(\mu \text{ or }\nu)$& $z$ & $1$ & $Z=\psi_z$ & $k_z^2$\\
			\midrule
			
			\multirow{3}{*}{Parabolic cyl.}
			& $u$ & $1$ & $U=\psi_u$ & $k_\perp^2u^2-\lambda$ \\
			& $v$ & $1$ & $V=\psi_v$ & $k_\perp^2v^2+\lambda$ \\
			& $z$ & $1$ & $Z=\psi_z$ & $k_z^2$ \\
			
			\bottomrule
		\end{tabularx}
		
		\caption{Liouville scaling factors for standard separable coordinate systems.
			Only coordinates with $s_i\neq1$ require rescaling.}
		\label{Tab1}
	\end{table}
	
	\begin{table}[h!]
		\centering
		\footnotesize
		\setlength{\tabcolsep}{3.5pt}
		\renewcommand{\arraystretch}{1.05}
		
		\begin{tabularx}{\columnwidth}{l c c c >{\raggedright\arraybackslash}X}
			\toprule
			\textbf{Ellipsoidal system} &
			\textbf{Axis parameters} &
			$\bm{s(q)}$ &
			\textbf{Liouville scaling} &
			\textbf{Effective $\Omega^2(q)$} \\
			\midrule
			Confocal quadrics
			& $q=\lambda,\mu,\nu$
			& $\sqrt{|(a^2-q)(b^2-q)(c^2-q)|}$
			& $X(q)=\psi(q)/\sqrt{s(q)}$
			& $Q_{SL}(q)\!-\!\frac{s''}{2s}\!+\!\frac{(s')^2}{4s^2}$ \\
			\bottomrule
		\end{tabularx}
		
		\caption{Master Liouville weight for the quadrics (ellipsoidal) family.
			Different coordinate systems correspond to parameter degenerations or interval choices for $q$; see \cite{MorseFeshbach}, Vol.~I, Chapter~5.}
		\label{Tab2}
	\end{table}
	
	\newpage
	
	\section{Example--Elliptical Coordinates}
	\label{AppendixC}
	\setcounter{equation}{0}
	\renewcommand{\thesection}{Appendix~\Alph{section}}
	\renewcommand{\theequation}{\Alph{section}.\arabic{equation}}
	
	\subsection*{Two--center Coulomb problem in 2D confocal elliptic coordinates}
\begin{figure}[ht]
		\centering
		\includegraphics[width=0.6\linewidth]{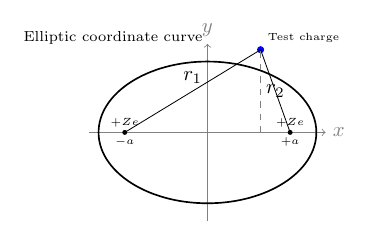}
		\caption{Geometry of the symmetric two--center Coulomb problem in two dimensions.
			The test charge is influenced by two identical charges $+Ze$ located at the foci
			$x=\pm a$ of an ellipse. The distances $r_1$ and $r_2$ enter naturally in confocal elliptic coordinates.}
		\label{fig:two_center_geometry}
	\end{figure}

	As a concrete illustration of the spatial EL invariant construction,
	we consider the motion of a test charge in a two–center Coulomb field in two
	dimensions (Fig.\ref{fig:two_center_geometry}). This system provides a genuinely non-trivial example: the potential
	is non-central, the Hamiltonian is separable only in confocal elliptic
	coordinates, and conventional quantum-mechanical treatments typically rely on
	perturbative, Born–Oppenheimer, or numerical methods.
	\vspace{0.5pc}
	
	\noindent In contrast, within the BM formulation, the stationary Schr\"{o}dinger
	equation admits an exact separation in elliptic coordinates, allowing the
	amplitude equations to be cast directly into Sturm–Liouville form and
	subsequently into EP equations. This enables the explicit
	construction of spatial EL invariants for both angular and radial degrees
	of freedom, providing an analytic characterization of stationary Bohmian
	dynamics without recourse to trajectory simulations.
	
	\begin{equation}
		\nabla^2 \Psi + \frac{2m}{\hbar^2}\bigl(E - V\bigr)\Psi = 0,
	\end{equation}
	with the symmetric potential
	\begin{equation}
		V(x,y) = -e^2\left(\frac{Z}{r_1} + \frac{Z}{r_2}\right),
		\qquad
		r_{1,2}=\sqrt{(x\mp a)^2+y^2}.
	\end{equation}
	This describes a charge moving in the field of two identical Coulomb centers separated
	by a distance $2a$ (no dipole moment).
	
	\subsubsection*{Elliptic coordinates and Laplacian}
	
	\noindent Introduce confocal elliptic coordinates $(\mu,\nu)$ via
	\begin{equation}
		x = a \cosh\mu \cos\nu, \qquad
		y = a \sinh\mu \sin\nu,
	\end{equation}
	with $\mu\ge 0$ and $\nu\in[0,2\pi)$.
	The distances to the foci satisfy
	\begin{equation}
		r_1 = a(\cosh\mu - \cos\nu), \qquad
		r_2 = a(\cosh\mu + \cos\nu).
	\end{equation}
	
	\noindent The scale factors are equal,
	\begin{equation}
		h_\mu = h_\nu = a\sqrt{\sinh^2\mu+\sin^2\nu}
		= a\sqrt{\cosh^2\mu-\cos^2\nu},
	\end{equation}
	and the Laplacian takes the diagonal form
	\begin{equation}
		\nabla^2 =
		\frac{1}{a^2(\sinh^2\mu+\sin^2\nu)}
		\left(
		\frac{\partial^2}{\partial\mu^2}
		+
		\frac{\partial^2}{\partial\nu^2}
		\right).
	\end{equation}
	
	\subsubsection*{Separation of variables}
	
	Writing $\Psi(\mu,\nu)=M(\mu)N(\nu)$ and defining
	\begin{equation}
		k^2=\frac{2mE}{\hbar^2}, \qquad
		\gamma=\frac{2me^2 a}{\hbar^2},
	\end{equation}
	multiplication of the Schr\"{o}dinger equation by
	$a^2(\sinh^2\mu+\sin^2\nu)=a^2(\cosh^2\mu-\cos^2\nu)$ yields
	\begin{equation}
		\Psi_{\mu\mu}+\Psi_{\nu\nu}	+
		\Bigl[a^2 k^2(\cosh^2\mu-\cos^2\nu)+2\gamma Z\cosh\mu \Bigr]\Psi=0.
	\end{equation}
	
	\noindent Separation introduces a constant $\Gamma$ and leads to
	\begin{align}
		M''(\mu) + \Omega_\mu^2(\mu) M(\mu) &= 0, \\
		N''(\nu) + \Omega_\nu^2(\nu) N(\nu) &= 0,
	\end{align}
	with
	\begin{align}
		\Omega_\mu^2(\mu) &= a^2 k^2\cosh^2\mu + 2\gamma Z\cosh\mu + \Gamma, \\
		\Omega_\nu^2(\nu) &= -a^2 k^2\cos^2\nu - \Gamma.
	\end{align}
	
	\noindent Both equations are already in self--adjoint Sturm--Liouville form with unit
	weight $s=1$ and require no Liouville rescaling.
	
	\subsection*{Angular Mathieu equation}

	Using $\cos^2\nu=\tfrac12(1+\cos2\nu)$, the angular equation becomes
	\begin{equation}
		N''(\nu)+\bigl(a_M-2q_M\cos2\nu\bigr)N(\nu)=0,
	\end{equation}
	with
	\begin{equation}
		a_M=-(\Gamma+\tfrac12 a^2k^2),
		\qquad
		q_M=\frac{a^2k^2}{4}.
	\end{equation}
	
	\noindent A fundamental independent solution pair is
	\[ y_1^{(\nu)}(\nu)=\mathrm{Ce}_\ell(\nu,q_M),\qquad
	y_2^{(\nu)}(\nu)=\mathrm{Se}_\ell(\nu,q_M).\]
	
	\noindent The associated EP amplitude is
	\begin{equation}
		\rho_\nu(\nu)	=
		\sqrt{ A_\nu\,y_1^{(\nu)2}	+	B_\nu\,y_2^{(\nu)2}	+	2D_\nu\,y_1^{(\nu)}y_2^{(\nu)}	},
	\end{equation}
	with constraint
	\begin{equation}
		A_\nu B_\nu - D_\nu^2 = \frac{k_\nu}{W_\nu^2},
	\end{equation}
	where $W_\nu=y_1^{(\nu)}{y_2^{(\nu)}}'-{y_1^{(\nu)}}'y_2^{(\nu)}$.
	
	\subsection*{Radial (hyperbolic) Mathieu equation}
	
	The radial equation may be viewed as the analytic continuation of the Mathieu
	equation to imaginary argument. A fundamental independent solution pair may be
	chosen as
	\[y_1^{(\mu)}(\mu)=\mathrm{Ce}_\ell(i\mu,q_M),\qquad
	y_2^{(\mu)}(\mu)=\mathrm{Se}_\ell(i\mu,q_M).\]
	
	\noindent The corresponding Ermakov amplitude is
	\begin{equation}
		\rho_\mu(\mu) =
		\sqrt{	A_\mu\,y_1^{(\mu)2}	+ B_\mu\,y_2^{(\mu)2} + 2D_\mu\,y_1^{(\mu)}y_2^{(\mu)}},
	\end{equation}
	with invariant constraint
	\begin{equation}
		A_\mu B_\mu - D_\mu^2 = \frac{k_\mu}{W_\mu^2},
	\end{equation}
	where $W_\mu$ is the constant Wronskian of the hyperbolic Mathieu pair.
	
	\paragraph{Remarks on Mathieu functions.}

	The Mathieu functions $\mathrm{Ce}_\ell(\nu,q_M)$ and $\mathrm{Se}_\ell(\nu,q_M)$
	form a complete set of even and odd solutions of the canonical Mathieu equation
	for real argument and fixed characteristic parameter $q_M$. Their analytic
	continuations to imaginary argument, $\mathrm{Ce}_\ell(i\mu,q_M)$ and
	$\mathrm{Se}_\ell(i\mu,q_M)$, provide corresponding independent solutions of the
	associated hyperbolic (modified) Mathieu equation. These functions and their
	properties—orthogonality, completeness, Fourier-series representations, and
	Wronskian structure—are treated in detail in classical references such as
	Morse and Feshbach \cite{MorseFeshbach} and Whittaker and Watson \cite{Whittaker}.
	Within the present framework, they serve as natural fundamental solutions for
	constructing EP amplitudes in elliptic coordinate systems.
	
	
\end{document}